\documentclass[sigconf,screen,nonacm]{acmart}

\usepackage[english]{babel}
\usepackage{bbding}
\usepackage{balance}
\usepackage{algorithm,algorithmic}
\usepackage{graphicx}
\usepackage{textcomp}
\usepackage[x11names,dvipsnames]{xcolor}
\hypersetup{hidelinks=true}
\usepackage{booktabs}
\usepackage{multirow}
\usepackage{scalerel}
\usepackage{tikz}
\usetikzlibrary{tikzmark}
\usetikzlibrary{positioning, shapes.symbols}
\usetikzlibrary{svg.path}
\usepackage{rotating}
\usepackage{tabularray}
\usepackage{xspace}
\usepackage{xcolor}
\definecolor{darkblue}{HTML}{0942A1}

\usepackage{cleveref}
\usepackage{soul}
\usepackage{subcaption}
\usepackage[flushleft]{threeparttable}
\usepackage{enumitem}
\usepackage{colortbl}
\usepackage[most]{tcolorbox}
\usepackage{mdframed}
\usepackage[bordercolor=white, backgroundcolor=white,prependcaption]{todonotes}
\usepackage{csquotes}
\usepackage{longtable} %

\usepackage{enumitem}
\setlist[itemize]{leftmargin=1em}

\usepackage{tikz}
\usetikzlibrary{shapes.geometric, arrows}

\tikzstyle{startstop} = [ellipse, rounded corners, minimum width=3cm, minimum height=1cm, text centered, draw=black]
\tikzstyle{process} = [rectangle, minimum width=3cm, minimum height=1cm, text centered, text width=3cm, draw=black]
\tikzstyle{decision} = [diamond, minimum width=2cm, minimum height=1cm, text centered, draw=black]

\tikzstyle{arrow} = [thick,->,>=stealth]

\usepackage{listings}

\tcbuselibrary{minted}
\usepackage{minted} % for html code display
        
\usepackage{mdframed}

\usepackage{tikz}
\usepackage{pgfplots}

\definecolor{lightgrey}{rgb}{0.95,0.95,0.95}

\newtcolorbox[auto counter, number within=section]{summary}[2][]{%
  colframe=blue!80!black, 
  colback=blue!10, 
  coltitle=black, 
  fonttitle=\bfseries, 
  title=Takeaway~\thetcbcounter: #2,#1,
  boxsep=1mm,   % Reduce spacing between text and border
  left=1mm,     % Adjust left padding
  right=1mm,    % Adjust right padding
  top=1.3mm,      % Adjust top padding
  bottom=1mm    % Adjust bottom padding
}

\crefname{section}{\S}{\S}
\crefname{subsection}{\S}{\S}
\crefname{subsubsection}{\S}{\S}

\newcommand{\plainLangDesc}{plain language description}
\newcommand{\plainLangDescCase}{Plain language description}
\newcommand{\PlainLangDesc}{Plain Language Description}

\newtheorem*{researchquestion*}{Research Question}

\author{Minela Bećirović}
\affiliation{%
  \institution{TU Braunschweig}
  \country{Germany}}
\email{mi.becirovic@tu-braunschweig.de}

\author{Ha Dao}
\affiliation{%
  \institution{Max Planck Institute for Informatics}
  \country{Germany}}
\email{hadao@mpi-inf.mpg.de}

\author{Mannat Kaur}
\affiliation{%
  \institution{Max Planck Institute for Informatics}
  \country{Germany}
}
\email{mkaur@mpi-inf.mpg.de}

\author{Martin Johns}
\affiliation{%
 \institution{TU Braunschweig}
 \country{Germany}}
\email{m.johns@tu-braunschweig.de}

\author{Alexandra Dirksen}
\affiliation{%
  \institution{University of Twente}
  \country{Netherlands}}
\email{a.dirksen@utwente.nl}


\begin{document}
%\listoftodos

% Cover Page
\pagestyle{empty}  % no headers/footers
\settopmatter{printacmref=false} % remove ACM reference block

\newpage

% Actual Anonymous Paper
\pagestyle{plain} % restore normal style

% \title{Tracking the Unseen: navigational tracking in YouTube Ad Clicks}

\title[Towards Simplified Cookie Banners using Plain Language]{Senseful Consense: Towards Simplified Cookie Banners\\using Plain Language}

\begin{abstract}

While the GDPR and ePrivacy Directive mandate that consent information must be clear and accessible, most modern cookie banners remain obscured by technical jargon, vague phrasing, and frequent content overload or underload.
This feasibility study investigates the impact of applying plain language (Einfache Sprache) to cookie banners within the IAB Transparency \& Consent Framework (TCF). 
In our study, we analysed cookie banner texts from 200 websites, using AI-based mapping to categorise extracted content into standardised processing purposes.

By substituting complex legal terms with simplified descriptions, we successfully demonstrated that the comprehension barrier can be lowered from a \textit{college-graduate level} to a \textit{7th-grade level}.
However, the effectiveness of plain language is inherently constrained by the informativeness of the original content; it cannot compensate for banners that omit legally required details. 
We conclude that while plain language is a vital tool for \textbf{digital accessibility}, it must be paired with \textbf{standardised implementation guidelines} to ensure that cookie banners are both readable and informative.
\end{abstract}

%\keywords{Plaintext}

\maketitle

\section{Introduction}
\label{sec:intro}
\begin{figure}[t!]
    \frame{\includegraphics[width=\linewidth]{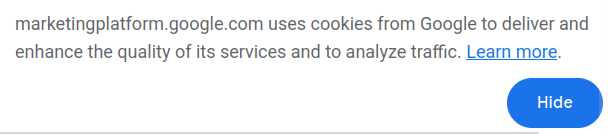}}
    \centering
    \caption{Example of a cookie banner that does not provide users with direct options to opt out or further customize their choices~\cite{GoogleMarketingPlatform}.}
    \label{fig:cb_ex_1}
\end{figure}

\begin{figure}[t!]
    \frame{\includegraphics[width=\linewidth]{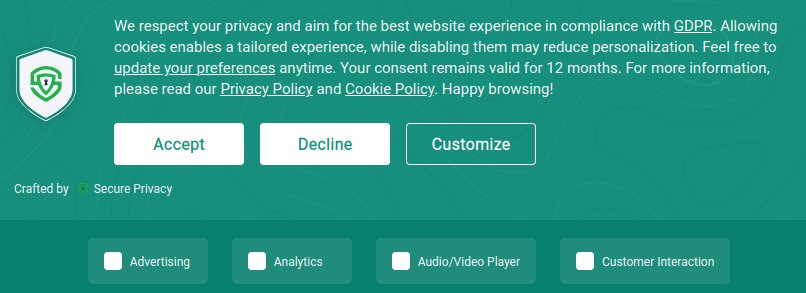}}
    \centering
    \caption{Example of a cookie banner that provides users with direct options to opt out and further customize their choices~\cite{SecurePrivacy}.}
    \label{fig:cb_ex_2}
\end{figure}

\noindent The General Data Protection Regulation (GDPR) states that the processing of personal data is only lawful under the fulfilment of certain conditions. For example, the processing of personal data is lawful if the consent of the data subject has been obtained for one or multiple purposes~\cite{GDPR}. This has led website operators to use cookie consent notices, also known as cookie banners, to inform users about the use of cookies on their websites and to obtain the necessary consent to process personal data~\cite{WVYP}. Article 7(2) of the GDPR requires that consent is obtained~\enquote{in an intelligible and easily accessible form, using clear and plain language}. The ePrivacy Directive (ePD) further supports this by requiring that users must be given~\enquote{clear and comprehensive information} about the purpose of the processing~\cite{ePrivacy}, so that users can make an informed decision. However, cookie banners differ not only in their design but also in the options they offer users regarding the extent of their consent. While some simply inform users about the use of cookies without offering further options, others offer detailed and sometimes very granular options, such as allowing users to exclude individual third-party services from the consent~\cite{UCSG}. Figures~\ref{fig:cb_ex_1} and~\ref{fig:cb_ex_2} show two different cookie banners, which differ significantly in their design and the options they offer.
\newline

% \subsection{Problem Statement and Motivation}
\noindent The European Data Protection Board (EDPB) suggests in their guidelines that the language used to obtain consent should be easily understood not only by experts but also by the average person~\cite{EDPB}. To ensure users can make an informed decision, the guidelines emphasise the use of clear, plain language. However, a survey conducted in 2024 found that three-quarters of participants did not understand the content of cookie banners~\cite{BR24}. This can be attributed to the fact that, unlike the requirements of the legal basis, cookie banners often use vague language and technical terms, which adds to the complexity of their content~\cite{CBLP}. Research investigating the effect of cookie banners on users also found that most users find them annoying rather than helpful~\cite{UPAR}. Dark patterns, which we discuss in Section~\ref{sec:darkPatterns}, increase this perception by attempting to influence the user's decision regarding the extent of the given consent~\cite{DPITW,forbrukerradet2018deceived}. 
Prior work on cookie banners has identified dark patterns ranging from overt manipulations of interface design and interaction~\cite{ontologydarkpatternsgray2024a,UCSG} to more subtle forms of \textit{textual} dark patterns, such as the use of overly complex language~\cite{kirkman2023darkdialogs}. 
In our work, we explore this particular \textit{type} of dark pattern by focusing on privacy-critical textual contexts --- such as cookie banner texts --- that, despite serving as users' first line of defence for making informed online data privacy decisions, are often not easily comprehensible.
% --- as a type of dark pattern, that has previously not been discussed in the community.
\newline
\newline
\noindent To address users' resulting frustration, tools such as browser extensions have been developed to minimise the need for user interaction with cookie banners~\cite{LSSOCB24}. Similar to ad blockers, they use a rule-based approach to first identify cookie banners on websites and then automatically apply the settings based on users' privacy preferences. Although this reduces the effort users have to put into interacting with cookie banners, it does not make the content of cookie banners more understandable to users. As a result, the problems of comprehensibility remain unresolved. Therefore, in this paper, we investigate the extent to which current cookie banner practices allow us to counter this problem. We argue that cookie banners should not only be understandable to experts - everyone should be able to easily understand their content. Rather than seeing cookie banners as a nuisance, users should perceive them as helpful tools that enable them to make informed decisions regarding their privacy.
\newline
\newline
In this paper, we conducted a feasibility study to investigate the extent to which cookie banners can be made more understandable and thus more accessible through the use of plain language, while considering the Interactive Advertising Bureau's Transparency \& Consent Framework (IAB TCF).
% Given these order-of-magnitude parameter changes, w
We ask the following question:
\begin{researchquestion*}
To what extent can plain language be used to improve the comprehensibility of cookie banners under the consideration of the IAB TCF within the current cookie banner landscape?
\end{researchquestion*}

Overall, our main contributions can be summarised as follows:
\begin{itemize}
\item We developed a set of~\textbf{plain-language descriptions} for each IAB TCF purpose to enhance the comprehensibility of cookie banner texts. Our evaluation demonstrates that these descriptions improve readability and user understanding (Section~\ref{sec:AddressingCompr} and~\ref{sec:eval}).
% \item We designed and implemented the \textbf{\browserEx{}} extension to automatically identify and extract the textual content of cookie banners, enabling large-scale analysis of current banner practices (see~\autoref{cookieBTX}).
% \item In this work we advocate to conceptualize privacy-critical textual context that is not easily comprehensible to the user to be considered as a dark pattern.
\item Using~\textbf{AI-based mapping} with OpenAI's o3 model, we mapped the extracted texts to the IAB TCF purposes, revealing inconsistencies and vague language in current cookie banner implementations (Section~\ref{methodology_catego}).
\item By mapping identified purposes to their corresponding plain language descriptions, we~\textbf{generated modified cookie banner texts} (Section~\ref{modifiedBannerText}).
\item We conducted a~\textbf{comparison of the original and modified cookie banner texts}, which showed improved comprehensibility while also highlighting that plain language cannot compensate for missing or unclear information (Section~\ref{modifiedBannerText}). We provide an open replication package containing the sampled websites and the full comprehensibility measurement results (before and after modification) for all 200 websites included in our study for further analysis\footnote{\url{https://anonymous.4open.science/r/ConSense-6ED0/README.md}}.
\end{itemize}

%Finally, we discuss... ethics, limitations in \autoref{sec:discussion}).
\section{Background \& Related Work}
\label{sec:background}
 % In this section, we give an overview of how the processing of personal data is regulated under the GDPR in the European Union (EU). First, we will define what personal data actually is and outline under what conditions it may be processed. For a better understanding of how this applies in the digital environment, we will explore cookies, cookie banners and briefly introduce the IAB TCF. We also provide an overview of prior work on cookie banners, focusing on user interaction, dark patterns and the development of browser extensions that minimize the interaction with cookie banners. Regarding plain language, we then provide a definition of plain language within the scope of this work.
 In this section, we outline how personal data processing is regulated under the GDPR in the EU. We define personal data, explain its processing conditions, and discuss cookies, cookie banners, and the IAB Transparency \& Consent Framework (TCF). We also summarise related work on cookie banners, including dark patterns and user interactions, and define plain language in the context of this paper.

% 2.1 GDPR (legal basis)
\subsection{The Legal Basis for Processing Personal Data} 
\label{sec:legalBasis}

The \textbf{General Data Protection Regulation} (GDPR) establishes the legal framework for data handling within the EU~\cite{GDPR}. 
Article 4\,(1) defines personal data broadly as any information relating to an ~\enquote{identified or identifiable natural person}, encompassing identifiers such as names or specific physical, cultural or social characteristics.

Under Article 4\,(2), ~\enquote{processing} is defined as any operation performed on this data, including its collection, storage, and disclosure.
While Article 6 outlines various conditions for lawful processing of this data, our work focuses on \textbf{consent}.
Article 4\,(11) defines consent as a ~\enquote{freely given, specific, informed and unambiguous} agreement by the data subject.
To ensure these criteria are met, Article 7\,(2) introduces the core requirement of our work: consent must be obtained in an ~\enquote{intelligible and easily accessible form}, utilising \textit{clear and plain language}.

\subsection{Cookie Banners and the IAB TCF} 
\label{sec:cookiesAndTCF}

To meet legal consent requirements, website operators utilise cookie banners to manage cookies, i.e., small data files stored on the user's device for session management, personalisation, or tracking, among others~\cite{mdnCookies,gdprCookies}.
Based on their function, these are categorised as (1)\,strictly necessary, (2)\,preference, (3)\,statistics, and (4)\,marketing cookies.
While the ePrivacy Directive\,(ePD)~\cite{ePrivacy} complements the GDPR by requiring explicit consent for all non-essential cookies, many modern website providers rely on Consent Management Platforms (CMP)~\cite{CMPcross,iab2025tcf}. These platforms, such as those implementing the IAB Transparency \& Consent Framework (TCF), provide a standardised mechanism for collecting and communicating user preferences across the digital advertising ecosystem~\cite{iabeurope2025tcfpolicies,UCSG}.
%
% use the IAB Transparency \& Consent Framework\,(TCF) to standardise this process across the digital advertising ecosystem.
%
%
% A cookie is a small data file stored on a user's device to manage sessions, personalize content, or track user behavior~\cite{mdnCookies,gdprCookies}. Based on their purpose, they can be categorized as: (1) strictly necessary, (2) preference, (3) statistics, and (4) marketing cookies. The latter types often involve processing personal data and therefore fall under the GDPR~\cite{gdprRecital30}. The ePrivacy Directive (ePD) complements the GDPR by requiring explicit user consent for any non-essential cookies~\cite{ePrivacy}. To meet these requirements, websites use cookie banners to inform users and obtain consent~\cite{WVYP,cookieEnforcer}. Many banners rely on Consent Management Platforms (CMPs), for example, those implementing the IAB's Transparency \& Consent Framework (TCF)~\cite{CMPcross,iab2025tcf}. The TCF provides a standardized mechanism for collecting, storing, and communicating user consent across the digital advertising ecosystem. It enables users to manage their preferences at a granular level, such as consenting to specific purposes or vendors~\cite{iabeurope2025tcfpolicies,UCSG}. This work focuses on TCF v2.2, in particular, its user-facing elements. The framework defines eleven purposes for data processing, allowing users to consent or object based on legal bases under GDPR Article 6(1)(a) (consent) and 6(1)(f) (legitimate interest):
The TCF provides a granular mechanism for users to manage preferences through eleven standardised processing purposes:

%\newline
\begin{minipage}[t]{0.98\linewidth}
\begin{enumerate}[label=\textbf{P\arabic*}, itemsep=0pt, topsep=.5pt, parsep=-1pt, partopsep=0pt, wide, labelwidth=!, labelindent=0pt]
\item Store and/or access information on a device
\item Use limited data to select advertising
\item Create profiles for personalised advertising
\item Use profiles to select personalised advertising
\item Create profiles to personalise content
\item Use profiles to select personalised content
\item Measure advertising performance
\item Measure content performance
\item Understand audiences through statistics or combined data
\item Develop and improve services
\item Use limited data to select content
\end{enumerate}
\end{minipage}
\newline
\newline
A more detailed list of the IAB TCF purposes can be found in Appendix~\ref{appendix_A}.
%\lex{define \textit{purposes} as IAB TCF purposes for data usage and make macro}
By utilising CMPs to implement this framework, operators allow users to consent or object based on legal bases such as GDPR Article\,6(1)(a) (consent) and 6(1)(f) (legitimate interest).
Earlier versions of the TCF were subject to legal scrutiny regarding transparency, legal basis, and data security~\cite{Maout2025, ICCL2025Tracking}. Although TCF\,v2.2 addressed many of these issues, the technical complexity inherent to these standardised descriptions remains a significant hurdle for user comprehension.

% 2.1.3. issues with consent banners
\subsection{The Crux with Cookie Banners}
\label{sec:issuesPriorResearch}

Despite the technical standardisation provided by frameworks like TCF, a significant gap remains between the legal requirement for transparency and actual user comprehension.

%In the following sections, we examine work on cookie banners, including dark patterns and user interactions.

% 2.1.3.1 dark patterns
\subsubsection{Dark Patterns in Banner Designs} \label{sec:darkPatterns}
Dark patterns are interface techniques that manipulate users into making decisions against their best interests~\cite{Gunawan_darkPat}, benefiting organisations --- such as website operators using cookie banners --- at the expense of users~\cite{Kollmer_darkPat}.
Various previous works have investigated this concept, how it manifests in the wild, and its effects on users.
Mathur et al. analysed dark patterns from interdisciplinary and normative perspectives, classifying various definitions and types\,\cite{whatmakesdarkmathur2021}. 
In 2024, Gray et al. describe dark patterns by the term \textit{deceptive design practices} and propose a multi-level ontology\,\cite{ontologydarkpatternsgray2024a}.
Utz et al. examined how cookie banner designs affect consent behaviour, identifying dark patterns such as highlighted ``accept'' buttons, hidden or hard-to-access options, and preselected choices~\cite{UCSG}. 
They found that these design elements, along with banner placement and option framing, significantly influence user decisions. 
The authors call for clearer consent guidelines to ensure that choices are informed and voluntary.
Krisam et al. analysed cookie banners on the top 500 German websites, finding that 86\% of those offering a ``reject'' option used visual nudging to encourage acceptance~\cite{DPITW}. 
The authors assume that rejecting cookies generally requires more effort on the part of users than accepting them. 
Based on these insights, they recommend raising user awareness of dark patterns and addressing their use through legal and privacy-protection measures.
% To the best of our knowledge we did not find any related work which considers textual content as dark patterns.
% In this work we advocate to conceptualize textual context that aims to deceive a user to be considered as a dark pattern.

% 2.1.3.2 comprehensibility issues
\subsubsection{Interaction of Users with Cookie Banners} \label{sec:intraction}

In 2023, Advances Metrics~\cite{AM} conducted a behavioural study on user interactions with cookie banners\footnote{Results should be interpreted with caution due to the lack of peer review.}. The CMP used allowed users to accept/reject all cookies, close the banner, or access detailed settings. The results showed that 68\% ignored or closed the banner, 25.4\% accepted all cookies, 5.3\% rejected all, and only 0.4\% accessed the detailed settings. A 2024 Bitkom Research survey~\cite{BR24} of about 1,000 internet users found that 76\% are annoyed by cookie banners. While 34\% consider cookie settings important, many admit that they don't understand them. Regarding behaviour, 24\% accept all cookies, 21\% reject all, and 33\% customise their choices. Rasaii et al.~\cite{exploringCookieverse} analysed cookie practices on the top 10,000 Tranco websites. Their automated tool effectively detected and interacted with cookie banners, though it sometimes misidentified unrelated elements due to banner diversity. Kulyk et al.~\cite{UPAR} explored users' perceptions of cookie disclaimers, finding generally negative attitudes driven by annoyance or privacy concerns. They noted users' limited understanding of cookies and suggested clearer explanations and simplified privacy policy summaries.

\subsection{The Concept of Plain Language} \label{sec:conceptPL}

To fulfil the transparency mandates established in Sec~\ref{sec:legalBasis}, institutional guidelines from the European Data Protection Board\,(EDPB) and the European Commission\,(EC) emphasise that information should be easily understood by the \enquote{average person} to build trust and enable democratic participation~\cite{EDPB}.
This shift toward functional clarity is reflected in a global movement toward standardised communication, exemplified by New Zealand's Plain Language Act~\cite{nzplain}, Canada's CAN-ASC-3.1:2025 – Plain Language standard~\cite{CNplain}, and Italy's UNI ISO 24495-1:2024 linguaggio chiaro standard~\cite{ITplain}.
Beyond clarifying government communication, Belgium and Canada also include recommendations for inclusive language (in Genderbewust taalgebruik~\cite{beplain2} and `Writing style and tone'~\cite{CNplain2}). 
These diverse efforts are largely anchored in the governing principles defined in ISO \textbf{24495-1:2023}, moving plain language from a stylistic choice to a legally significant component of digital accessibility~\cite{ISO24495-1_eng}.
The number of regulations governing the implementation of plain language underscores the importance of considering its legal binding.

\subsubsection{Defining the Model: Einfache Sprache (Germany)}

Although most of the aforementioned concepts aim to simplify complex texts, their target audiences and linguistic implementations should be considered, as they may stem from different use cases. 
In the context of our study, a critical distinction is made between two simplification concepts, usually used in Germany, that are often confused: \enquote{Leichte Sprache}and \enquote{Einfache Sprache}~\cite{lbit_leichte_sprache}.
\begin{description}
    \item[Leichte Sprache] is a strictly regulated system designed for individuals with cognitive impairments; it utilises short sentences and a highly reduced vocabulary that, e.g., forbids technical terms~\cite{bmfsf_barrierefreiheit_leichte_sprache}.
    \item[Einfache Sprache] targets a broad, general audience. It allows for the inclusion of technical terms while simplifying complex syntax and grammar to ensure the full content of the original text remains intact~\cite{grundbildung_leichte_sprache}.
\end{description}

\noindent Einfache Sprache can therefore be seen as an extension of Leichte Sprache or as a reduction of standard language~\cite{grundbildung_leichte_sprache}.
Our study adopts \enquote{Einfache Sprache} as its operational definition of \textit{plain language}, specifically targeting the language levels of B1-B2.

\begin{definition}[Plain language]
    In our work, we use the concept of Einfache Sprache as a role model for simplifying cookie banner content. Whenever \textit{plain language} is mentioned, it refers specifically to the concept of Einfache Sprache (as defined in DIN ISO 24495-1~\cite{DINISO24495-1} and DIN 8581-1~\cite{DIN8581-1}).
\end{definition}

%2.4.2 The scope of plain language standards - para that describes which \textit{type} of text is subject to plain language requirements and also how cookie banners fit under this scope.
%\subsubsection{The Scope of Plain Language Standards} Plain language laws, standards, and guides are applicable to public-facing information and documents, which include legal texts such as data privacy policies. Limited prior work has examined how plain language is being implemented in public organizations~\cite{giacomin2024plain}, in the legislative arena~\cite{arenas2023publish}, in interactive system design~\cite{oliveira2025plain}, and devised methods for evaluating the use of plain language in public information portals~\cite{nunes2023method, oliveira2023assessment}.
% \paragraph{Cookie banners as legal texts}
%Given that cookie banners are meant to convey information required by privacy and data protection regulations and serve as a way of obtaining \textit{informed} consent, they can be conceptualized as legal texts (and have been examined as such~\cite{UCSG, nouwens2020dark, CBLP}) to which the requirements of plain language apply.
% Exiting research has examined them as such, evaluating them against legal frameworks~\cite{UCSG, nouwens2020dark, CBLP}.
\subsubsection{The Scope of Plain Language} Plain language guidelines are generally intended for public-facing information and documents, which include legal texts such as data privacy policies. Limited prior work has examined how plain language is being implemented in public organizations~\cite{giacomin2024plain}, in the legislative arena~\cite{arenas2023publish}, in interactive system design~\cite{oliveira2025plain}, and devised methods for evaluating the use of plain language in public information portals~\cite{nunes2023method, oliveira2023assessment}. Given that cookie banners are intended to convey information required by privacy and data protection regulations and serve as a way of obtaining \textit{informed} consent, they can be conceptualised as legal texts (and have been examined as such~\cite{UCSG, nouwens2020dark, CBLP}) for which plain language principles are highly relevant.

% \mk{check for govt-related research studies on making govt websites more understandable.}

% \newline
% \newline
% \noindent For example, the ORF.at website introduced a module in 2020 that makes the news accessible to readers regardless of their language level by presenting the most important news of the day in Einfache Sprache~\cite{orf_einfache_sprache_2020}. The feature can be accessed by navigating to the Einfache Sprache section of the website, which is shown in Figure~\ref{fig:orf_ES}. Once activated, the readers are provided with a selection of the day's most important news articles written in Einfache Sprache.

% \begin{figure}[!h]
%     \centering
%     \frame{\includegraphics[width=0.67\textwidth]{media/06_ORF.png}}
%     \caption[Interface element to enable the most important news of the day in Einfache Sprache.]{Interface element to enable the most important news of the day in Einfache Sprache~\cite{ORF}.}
%     \label{fig:orf_ES}
% \end{figure}

% 2.2.2. Translation tools
% 2.4.3 Tools of plain language - introducing and explaining capital.ai and Wortlinga.
\subsubsection{Tools for Plain Language}
\label{toolsPL}
Several tools can rephrase text in plain language. Within the scope of our work, we used the plain language functionality provided by capito.ai's and Wortliga's translation tools. The latter is a tool recommended by Germany's certification authorities~\cite{neueskitoolmachtlingscheid2024}. 
The efficacy of both tools has been compared and reported by Ludewig and Reich, and Manning~\cite{kitoolsundleichtereich2025, kitoolsfuereinfachemultisprech2023}.

The tool by capito.ai offers several features, including checking the comprehensibility of a text, supporting gender-sensitive language, and simplifying text into three language levels: A1-A2, A2-B1, and B1-B2~\cite{capito, capito_faq}. The AI-based tool has been trained on more than 6,000 translation projects, with its learning process still ongoing. For analysis or simplification, the text can be copied into the tool. It then assesses the comprehensibility of the text and determines to what extent it matches each of the three language levels: A1-A2, A2-B1, and B1-B2. For each language level, a percentage is provided indicating how strongly the analyzed texts can be considered to be written in that level~\cite{capitoAI}. The concept of plain language that was defined in Section~\ref{sec:conceptPL} is reflected by the language level B1-B2, as stated by capito.ai. A more detailed description of each language level is provided in Appendix~\ref{appendix_B}. To simplify the text, one language level must be chosen directly as the target language level. Four languages are currently fully supported: German, English, French, and Spanish. Regarding data protection, it is stated that texts are handled with care and not shared with third parties. %Additionally, they offer expert services for translating texts into plain language and provide workshops and training courses for it. 

Wortliga offers two features designed to assist with text rewriting using AI technology: the Wortliga Rewriter and the Wortliga Text Analysis. 
For example, using AI-powered features, users can rephrase text into plain language in over 50 languages~\cite{wortliga2024, wortliga2024plainlanguage}. %The basic version of Wortliga can be used for free without the need for registration. However, the plain language function is only included in the premium version.
Wortliga values data protection and security, meaning that the texts are handled in accordance with the GDPR and are not stored or used for AI training. For analysis or simplification, the text simply needs to be copied into the chosen feature.

%\hd{How can we believe in the output of Wordliga and capito.ai?, any related work used these tools?}
%\minb{Just some notes: captio.ai (https://www.capito.eu/en/faqs/): expert for Easy Language and accessible information for over 20 years; based on an artificial intelligence; provided with over 6,000 translation projects from the capito network to learn what is easy or difficult to understand; TÜV-certified (https://www.capito.eu/en/capito-method/) Wortliga (https://wortliga.de/): AI technology where the language models are continously upaded; recommended by reputable institutions such as TÜV Rheinland (external validation that suggests that professional organizations consider their results dependable); relies on the Hamburger Verständlichkeitskonzept for text clairty }

% 2.3. Assessment of textual comprehensibility
\subsection{Assessment of Textual Comprehensibility}
\label{sec:assesmentCompr}

% 2.3.1. scores for assessment
% 2.3.2. tools for assessment

Several methods exist to assess the comprehensibility of text, such as the Flesch Reading Ease~\cite{derivationnewreadabilitynotes} and the New Dale-Chall Formula~\cite{dale1948formula}. 
Many linguistic tools implement these methods and can be used to assess the understandability of a text by computing readability scores and language levels.

%To assess the comprehensibility of a text, user perceptions are typically measured through surveys. However, this would require a more in-depth investigation drawing on expertise in usability and accessibility. As our goal is to assess our proposal's potential, we decided to rely on established tools for the measurement process. We use the Python library Textstat and the text analysis features provided by capito.ai and Wortliga. These tools compute statistics, such as readability scores and language levels, for each analyzed text.
One such tool is the Python library Textstat that computes statistics from text, enabling us to assess factors such as readability, complexity, and the grade level required to understand the analysed text~\cite{textstat}. Although the library offers a wide range of functions, this paper considers only the Flesch Reading Ease Formula, the Gunning FOG Formula, and the New Dale-Chall Formula. 
The Flesch Reading Ease Formula computes the Flesch Reading Ease Score, which indicates how easy a text is to read; higher scores indicate improved readability. 
The Gunning FOG Formula computes the FOG Index of a text, which reflects the grade level at which a reader should be able to understand the text. 
Therefore, a lower score indicates improved readability. 
The New Dale-Chall Formula uses a look-up table of the 3,000 most commonly used English words to compute the Dale-Chall Readability Score, which also reflects the grade level at which a user should be able to understand the text. A lower score indicates improved readability. 

Other exemplary tools are provided by capito.ai and Wortliga, as introduced in Section~\ref{toolsPL}. 
They implement text analysis features that allow determining the language level of the analysed texts. 
For capito.ai, we consider only the percentage computed at the language level B1-B2, as it reflects the concept of plain language. 
The Wortliga Text Analysis feature also provides language-level information and computes a comprehensibility score. If the score is between 60 and 100, the text is considered to be comprehensible. 
The Wortliga Rewriter also computes a readability score (RS), with 100\% indicating the best possible readability.

%It is important to note that we do not aim to prove the concept's effectiveness in terms of comprehensibility. Instead, we want to provide an initial assessment of its potential to improve the comprehensibility for users. The actual effectiveness of the concept in regard to the comprehensibility is influenced by additional factors that are beyond the scope of this work (see Section~\ref{sec:limitations}).

% \subsubsection{Dark Patterns in Banner Texts}
\subsection{Research Gaps}
% \mk{maybe include related dark patterns work here?}
The implementation of dark patterns has already been investigated across various contexts, for example, within UX Design~\cite{darkpatternssidegray2018a}.
In the context of cookie banners, previous research has primarily focused on the legal validity of their textual content\,\cite{automaticclassificationlegalvanhofslot2022} or the effect of dark patterns in design elements on users' perceptions~\cite{borberg2022so} and consent decisions~\cite{bielova2024effect}, as discussed in Sec.\,\ref{sec:darkPatterns}.
More recent work has also considered complex language when detecting various dark patterns on cookie dialogs~\cite{kirkman2023darkdialogs}. 
% However, to the best of our knowledge, there exists no related work that considers the complexity of the coockie banner content. 
% of cookie banners as a form of dark pattern.
%\lex{complexity of text - used against people. Adapt the paragraph after!}
However, there is lack of research on implementing concepts such as plain language in cookie banners, underscoring an underexplored aspect of cookie banners. 
Our work aims to address this gap by investigating how cookie banners can be made more understandable and accessible through plain language, while considering the IAB TCF.
\section{Comprehensibility of Cookie Banners}
\label{sec:AddressingCompr}
In the following sections, we reflect on the IAB TCF's efforts to make their purposes understandable to users and explore how plain language can be used to enhance these efforts.

% 3.1. Do TCF's user friendly descriptions (UFD) of purposes tackle 2.1.3.2.?
% 3.1.1. assessing comprehensibility using 2.3
% 3.2. plain language for the win?
% 3.2.1. translating UFD into plain language descriptions (PLD)
% title just as a placeholder
\subsection{Enhancing TCF's Efforts}
Each of the IAB TCF purposes introduced in Section~\ref{sec:cookiesAndTCF} includes an additional description of their purposes, which they refer to explicitly as ``user-friendly description'', as shown in Appendix~\ref{appendix_A}. When applying the tools for the assessment of the comprehensibility mentioned in Section~\ref{sec:assesmentCompr} to the TCF's own user-friendly descriptions, we observe poor comprehensibility scores (see Table~\ref{tab:langlevels-readscore}, column~\enquote{TCF Purposes (User-Friendly Description)}). 

To explore the extent to which plain language, as defined in Section~\ref{sec:conceptPL}, can provide users with a better understanding of what they are consenting to regarding their personal data, we apply it to the user-friendly descriptions of the IAB TCF purposes.

For the rephrasing, we use the plain language function provided by the Wortliga Rewriter. During the rephrasing process, we use the language level and the readability score provided by the Rewriter to track any improvements in readability. We generate various versions until the language level is suitable and the readability score shows no significant improvement. We consider a language level to be suitable if it ranges from B1-B2, as this range reflects the concept of plain language as explained in Section~\ref{toolsPL}. In some cases, we also used capito.ai for rephrasing or manually adjusted the language level or readability score (e.g., changing \enquote{advertisement} to \enquote{ad}). The version of the rephrased text that has a suitable language level and shows no significant improvement in its readability score when changed is used as the~\plainLangDesc{} for the purpose. 
Table~\ref{tab:explanation-texts} shows the~\plainLangDesc{} for 11 purposes, its label, and the generated~\plainLangDesc{}.

\begin{table*}[!h]
\small
     \centering
    
    \resizebox{\textwidth}{!}{
    \begin{tabular}
    {l|p{0.25\textwidth}|p{0.7\textwidth}}
    \toprule
    \textbf{Label} & \textbf{Purpose} & \textbf{\PlainLangDesc{}} \\
    \midrule
    
    P1 & Store and/or access information on a device & We store or access small pieces of data, which are called cookies, or similar online identifiers on your device. We also gather other details, such as your browser type and language settings. This allows us to recognize your device whenever it connects to an app or website. We do this for one or more purposes.\\
    \hline
    P2 & Use limited data to select advertising & Ads you see on this service might use limited information. This information includes the website or app you are currently using. We also use your general location, but it is not exact. We check the type of device you use, like a phone or tablet. Additionally, we look at what content you are viewing or have viewed before. This helps us decide which ads to show, so you don't see the same ad too many times.\\
    \hline
    P3 & Create profiles for per\-son\-al\-ised advertising & We store information about your actions on this service. This includes forms you fill out and content you view. We can combine this with other details about you. For example, we might use data from your past activities. This includes your activity on this service, other websites, or apps. We might also use data from users who are similar to you. We use this combined information to create or improve a profile about you. This profile may show your interests and personal details. Later, we or other companies can use your profile to show you ads that match your possible interests.\\
    \hline
    P4 & Use profiles to select personalised advertising & We show you ads here. These ads are based on your ad profiles, which indicate your activity on this service. They also include what you do on other websites or apps, for example, the forms you fill out. This also includes the content you view online. These profiles also take into account your potential interests. They also incorporate some personal information.\\
    \hline
    P5 & Create profiles to personalise content & We store information about your actions on this service. This includes forms you fill out and content you view. We can combine this with other details about you. For example, we might use data from your past activities. This includes your activity on this service, other websites, or apps. We might also use data from users who are similar to you. We use this combined information to create or improve a profile about you. This profile may show your interests and personal details. Later, we use this profile to show you content that we think you will like. For instance, we might change the order of what you see to make it easier for you to find things that fit your interests.\\
    \hline
    P6 & Use profiles to select personalised content & We show you content on this service that is based on your personalized profile. This profile, which we create from your activity, reflects what you do on this and other services. It includes the forms you fill out and the things you view. We also consider your possible interests and personal details. For example, we might change the order of the content you see. This makes it easier for you to find non-advertising content that fits your interests.\\
    \hline
    P7 & Measure advertising performance & We use information about the ads you see and how you interact with them. For example, we check if you saw an ad or clicked on it. We also see if it made you buy something or visit a website. This helps us understand if an ad worked well for you or others. We can then see if the ad achieved its goals. This information is very useful for us to see how relevant advertising campaigns are.\\
    \hline
    P8 & Measure content performance & We use information about the content you see and how you interact with it. This helps us understand if the non-advertising content reached the right audience. We also want to know if it matched your interests. For example, we look at whether you read an article or watch a video. We also consider if you listen to a podcast or view a product description. We also track how long you spend using a service and which websites you visit. This helps us understand how relevant the non-advertising content is that we show to you.\\
    \hline
    P9 & Understand au\-di\-en\-ces through sta\-tis\-tics or com\-bi\-na\-tions of data from different sources & We create reports by combining different types of information. This information includes user profiles, sta\-tis\-tics, market research and data from analytics. We look at how you and other users interact with ads and non-advertising content. By doing this, we can find shared traits among users. For example, we can see which groups of people are more interested in an ad campaign or specific content.\\
    \hline
    P10 & Develop and improve services & We use information about how you use this service, such as when you click on ads or content. This helps us make our products and services better. We also use it to create new products and services. For example, we consider how users interact with our service and what kind of audience we have. We do not use this information to create or improve user profiles and identifiers.\\
    \hline
    P11 & Use limited data to select content & Content you see on this service might use limited information. This information includes the website or app you are currently using. We also use your general location, but it is not exact. We check the type of device you use, like a phone or tablet. Additionally, we look at what content you are viewing or have viewed before. This helps us decide which content to show, so you don't see the same video or article too many times.\\
    \bottomrule
    \end{tabular}
    }
    \caption{~\plainLangDescCase{}s for the purposes written in plain language.} \label{tab:explanation-texts} 
\end{table*}

% Table~\ref{tab:explanation-texts} shows the~\plainLangDesc{} for the first purpose of the IAB TCF. The full table listing each purpose, its label, and the generated~\plainLangDesc{} can be found in Appendix~\ref{appendix_C}.

% \begin{table}
%      \centering
%     \resizebox{\linewidth}{!}{
%     \begin{tabular}
%     {l|p{0.15\textwidth}|p{0.45\textwidth}}
%     \toprule
%     \textbf{Label} & \textbf{Purpose} & \textbf{\PlainLangDesc{}} \\
%     \midrule
    
%     P1 & Store and/or access information on a device & We store or access small pieces of data, which are called cookies, or similar online identifiers on your device. We also gather other details, such as your browser type and language settings. This allows us to recognize your device whenever it connects to an app or website. We do this for one or more purposes.\\
%     \hline
%     \end{tabular}
%     }
%     \caption{~\plainLangDescCase{}s for Purpose P1 written in plain language.} \label{tab:explanation-texts-P1} 
% \end{table}
% \vspace*{-1cm}
Table~\ref{tab:langlevels-readscore} reports the changes in the language levels and readability scores observed during the rephrasing process for both the original text (i.e., the user-friendly description) and the rephrased text (i.e., the~\plainLangDesc{}).
A score of 100\% indicates the best readability possible; Readability decreases as values move towards the negative range.

\begin{table}[!h]
    \centering
    \resizebox{\linewidth}{!}{
    \begin{tabular}{lcccc}
        \toprule
        \multirow{2}{*}{Purpose Label} & 
        \multicolumn{2}{c}{\shortstack{TCF Purpose \\ (User-friendly Description)}} & 
        \multicolumn{2}{c}{\shortstack{Rephrased Text \\ (\PlainLangDesc{})}} \\
        \cmidrule(lr){2-3} \cmidrule(lr){4-5}
        & LL & RS & LL & RS \\
        \midrule
        P1 & C2 & -97\% & B1 & 70\% \\
        P2 & C1 & -58\% & B1 & 75\% \\
        P3 & C2 & 14\%  & B1 & 77\% \\
        P4 & C2 & -24\% & B1 & 79\% \\
        P5 & C2 & -11\% & B1 & 76\% \\
        P6 & C2 & -8\%  & B1 & 74\% \\
        P7 & C2 & 33\%  & B1 & 74\% \\
        P8 & C2 & 22\%  & B1 & 72\% \\
        P9 & C2 & -77\% & B1 & 68\% \\
        P10 & C2 & 18\% & B1 & 72\% \\
        P11 & C2 & -68\% & B1 & 74\% \\
        \bottomrule
    \end{tabular}
    }
    \caption{Comparison of the language levels (LL) and readability scores (RS) of TCF's original user-friendly description and our new rephrased descriptions in plain language.}
    \label{tab:langlevels-readscore}
\end{table}

% 3.2.2. assessing comprehensibility of PLD & comparing to UFD
% title just as a placeholder
\subsection{Enhancement Comparison}
\label{enhancementComp}
In order to assess the comprehensibility of the user-friendly descriptions and the~\plainLangDesc{}s from various perspectives, we decided to utilise various tools introduced in Section~\ref{sec:assesmentCompr}, namely, the library Textstat and the text analysis features provided by capito.ai and Wortliga.
The results of the comparison are presented in Table~\ref{tab:UD-cmprocess}.

\begin{table}[htbp]
    \centering
    
    \setlength{\tabcolsep}{4pt}
    \renewcommand{\arraystretch}{1.15}
    \resizebox{\linewidth}{!}{
    \begin{tabular}{lcccccccccccc}
    \toprule
    & \multicolumn{2}{c}{\textbf{\shortstack{Flesch Reading \\ Ease Score}}} 
    & \multicolumn{2}{c}{\textbf{\shortstack{FOG \\ Index}}} 
    & \multicolumn{2}{c}{\textbf{\shortstack{Dale–Chall \\ RS}}} 
    & \multicolumn{2}{c}{\textbf{\shortstack{capito.ai \\ B1–B2}}} 
    & \multicolumn{2}{c}{\textbf{\shortstack{Wortliga \\ Lang. Level}}} 
    & \multicolumn{2}{c}{\textbf{\shortstack{Wortliga \\ CS}}} \\
    \cmidrule(lr){2-3}\cmidrule(lr){4-5}\cmidrule(lr){6-7}\cmidrule(lr){8-9}\cmidrule(lr){10-11}\cmidrule(lr){12-13}
    \textbf{Purpose} 
    & {UFD} & {PLD}
    & {UFD} & {PLD}
    & {UFD} & {PLD}
    & {UFD} & {PLD}
    & {UFD} & {PLD}
    & {UFD} & {PLD} \\
    \midrule
    
    P1 & 30.57 & 65.93 & 12.68 & 8.43 & 10.61 & 9.49 & 0\% & 87\% & C2 & B1 & 0 & 93 \\
    P2 & 27.16 & 75.5 & 25.42 & 7.71 & 11.99 & 8.01 & 0\% & 92\% & C2 & B1 & 3 & 92 \\
    P3 & 33.28 & 68.57 & 15.81 & 7.25 & 9.4 & 7.79 & 30\% & 99\% & C2 & B1 & 32 & 93 \\
    P4 & 29.86 & 69.58 & 20.3 & 8.67 & 11.06 & 9.06 & 2\% & 96\% & C2 & B1 & 9 & 96 \\
    P5 & 33.62 & 76.01 & 18.49 & 7.18 & 9.58 & 7.35 & 0\% & 97\% & C2 & B1 & 27 & 85 \\
    P6 & 34.43 & 66.74 & 18.98 & 9.3 & 9.99 & 8.33 & 0\% & 81\% & C2 & B1 & 25 & 91 \\
    P7 & 54.56 & 75.4 & 14.27 & 7.72 & 9.09 & 7.35 & 51\% & 98\% & C2 & B1 & 27 & 94 \\
    P8 & 50.67 & 66.33 & 14.54 & 9.62 & 9.66 & 7.83 & 1\% & 98\% & C2 & B1 & 20 & 91 \\
    P9 & -17.18 & 50.02 & 30.49 & 10.12 & 14.5 & 10.69 & 2\% & 97\% & C2 & B2 & 0 & 91 \\
    P10 & 26.64 & 57.67 & 19.0 & 9.56 & 10.74 & 7.79 & 95\% & 99\% & C2 & B1 & 0 & 87 \\
    P11 & 24.11 & 66.74 & 27.09 & 8.79 & 11.82 & 7.93 & 0\% & 92\% & C2 & B1 & 0 & 86 \\
    \bottomrule
    \end{tabular}
    }
     \caption{Results of the comprehensibility measurements for the user-friendly descriptions (UFD) and the~\plainLangDesc{}s (PLD). Abbreviations: RS = Readability Score, Lang. Level = Language Level, CS = Comprehensibility Score.}
     \label{tab:UD-cmprocess}
\end{table}

For the Flesch Reading Ease Score, the~\plainLangDesc{}s received higher scores than the user-friendly descriptions, indicating an improvement in readability. In particular, the user-friendly description of purpose P9 stands out as it received a score of -17.18, while the corresponding~\plainLangDesc{} received a score of 50.02. For the FOG Index, Table~\ref{tab:UD-cmprocess} shows that the~\plainLangDesc{}s received lower scores than the user-friendly descriptions, indicating improved readability. The user-friendly description of purpose P3 received a score of 15.81, which means that a college junior should be able to understand it. The corresponding~\plainLangDesc{} received a score of 7.25, indicating a seventh grader should be able to understand the text. We observe the same pattern for the Dale-Chall Readability Score, as the~\plainLangDesc{}s consistently received lower scores than the user-friendly descriptions, indicating an improvement in readability. The user-friendly description of purpose P5 received a score of 9.58, which means that a 13th- to 15th-grade (college) student should be able to understand it. The corresponding~\plainLangDesc{} received a score of 7.35, indicating that a 9th- or 10th-grade student should be able to understand the text.

The percentages that were computed by capito.ai for the language level B1–B2, presented in Table~\ref{tab:UD-cmprocess}, are consistently higher for the~\plainLangDesc{}s than for the user-friendly descriptions. For example, the percentage increased from 0\% to 97\% for purpose P5. Similarly, the language levels determined by the Wortliga text analysis tool also show consistent improvement. The user-friendly descriptions received a C2 level, while the corresponding~\plainLangDesc{}s ranged from B1 to B2. The comprehensibility scores computed by the Wortliga text analysis tool show that the user-friendly descriptions received scores between 0 and 32, while the~\plainLangDesc{}s received scores between 85 and 96. Therefore, the~\plainLangDesc{}s are considered comprehensible, while the user-friendly descriptions are not.

% 3.3. Let's apply PLD on consent banners then!
\subsection{Application of~\PlainLangDesc{}s}

Based on these results, we were motivated to explore the extent to which the use of plain-language descriptions in cookie banners could also improve comprehensibility with respect to our research question. Therefore, in this work, we will examine the feasibility of a concept designed to improve the comprehensibility of cookie banners. This concept combines the purposes defined in the IAB TCF (see Section~\ref{sec:cookiesAndTCF}) with the principles of plain language as defined in Section~\ref{sec:conceptPL}.
% 4. Methodology
\section{Methodology}
\label{sec:method}
In the following sections, we present our methodological approach for examining the extent to which plain language can improve the comprehensibility of cookie banner texts in relation to our research question.

% 4.1. Overview
\subsection{Study Design}
\begin{figure}[!h]
    \begin{tikzpicture}[
  block/.style={
    signal,
    text width=2.27cm,
    font=\fontsize{7}{10}\selectfont,
    signal pointer angle=140,
    minimum height=1.27cm,
    text centered,
    inner sep=4pt,
    text=white
  }
]
  \node[block, draw=darkblue, fill=darkblue!70] (a)
    {Mapping\\based on the\\IAB TCF Purposes};

  \node[block, draw=darkblue, fill=darkblue!70, right=0.2mm of a, signal from=west] (b)
    {Creation of Modified\\Cookie Banner Texts};

  \node[block, draw=darkblue, fill=darkblue!70, right=0.2mm of b, signal from=west] (c)
    {Measurement of\\Comprehensibility};

\end{tikzpicture}
    \caption{The three main components of the methodology.}
    \label{fig:mainCompDiag}
\end{figure}
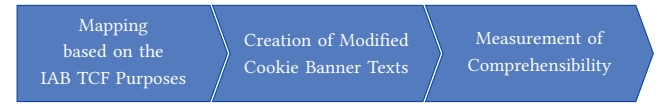

To obtain a representative overview of current cookie banner practices, a sample of 200 websites with English-language cookie banners will be analysed. For each cookie banner, the analysis follows a four-step process: First, the text from the first level of the cookie banner is extracted. It is then mapped to the IAB TCF purposes. Based on this, a modified version of the banner text, written in plain language, is created. Finally, the comprehensibility of both the original and modified cookie banner texts is assessed. 
This was done using the comprehensibility assessment approach described in Section~\ref{sec:assesmentCompr}. Within the scope of this paper, we focus on the three main components illustrated in Figure~\ref{fig:mainCompDiag}, which are described in more detail in the following sections.

% 4.2. Preliminaries
\subsection{Data Collection}

%\mj{I interpret the term 'preliminaries' with 'required knowledge'. Something in the line of 'data collection and preprocessing' appears to be a better fitting title}

This section outlines components of our approach that serve as preparatory steps for comprehensibility assessment.

% 4.2.1. Website samples
\subsubsection{Selection of the Website Sample} \label{selectionOfWS}
%\hd{Todo: Responsible}

%\hd{why do we think 200 is enough?\\ - we put our focus on parsing textual content, not feasibility of banner detection! \\- we decided against doing  invasive web crawling, see. Sec. \ref{ethix}.}

To obtain a representative overview of current cookie banner practices, we used the Majestic Million list~\cite{majestic_million,majestic_glossary_million}, which ranks the top 1,000,000 websites by the number of referring subnets. The list provides broad coverage of different websites, ensuring our study includes a diverse set of examples rather than being limited to a single type. To select 200 websites with English-language cookie banners from the full list, we conducted a manual review by sequentially processing the list. For each website, we checked whether a cookie banner was displayed and if the banner text was written in English. To increase the likelihood that the banner text would be displayed in English, we set the browser language to English in the browser settings. If the banner was still not displayed in English, the language was manually adjusted in the website settings, if offered. No translation tools were used to convert the banner text to English, as we considered only banners that were natively available in English on the website. In addition, to ensure that only one version of a website was included, we removed similar versions of the same website. For example, if~\texttt{google.com} was selected,~\texttt{google.de} was excluded. Similarly, a subdomain such as~\texttt{play.google.com} was excluded if it displayed a cookie banner identical to that of its root domain,~\texttt{google.com}. 
This approach was taken to avoid redundancy and increase the diversity of cookie banner implementations in the website sample. Websites that were unavailable during the manual review or that did not display a cookie banner at all were excluded. The filtering process continued until a final set of 200 qualifying websites was obtained. A table containing the full website sample is available in the released artefact.
% Appendix~\ref{appendix_D}.

\subsubsection{Cookie Banner Text Extraction}
\label{cookieBTX}

To identify and extract the text from cookie banners, we created a Firefox browser extension. Various methods for automatically detecting cookie banners on websites have already been developed. The approaches range from using filter lists containing CSS selectors~\cite{impactCookieNotices}, traversing the HTML tree~\cite{soe2022automateddetectiondarkpatterns}, to using BERT classifiers~\cite{cookieEnforcer}. However, the aim of this work is not to create a market-ready solution for identifying or extracting text from cookie banners. 
Instead, we developed a tool to solely collect the data needed to assess the feasibility of our approach. 
A 2023 study found that only 0.4\% of participants viewed the detailed options presented at subsequent levels of cookie banners~\cite{AM}. 
To match the user's reality, we decided to analyse only the first level of the cookie banner, which most users actually interact with. 
Because cookie banners are implemented in various ways, automatic detection is challenging. The extension prompts the user to manually click the banner to identify it and trigger the extraction process. Using our extension, we extracted the cookie banner texts from 181 of the 200 websites in our sample. For the remaining 19 websites, extraction failed, and the extension displayed an error. 
Among these websites, most implemented their cookie banners using Shadow DOM, whereas a few used iFrames. 
It was also observed that, in some cases, the extension could only extract parts of the banner text. Certain visually hidden elements, such as dynamically generated or customised lists, remained invisible during extraction despite attempts to render them visible. 
These banner texts were not discarded, as they represent what users actually see before interacting with the banner.

% 4.3. Mapping banner texts to TCF purposes (briefly describe why o3 is used-> ref to Table 4 in Apx)
\subsection{IAB TCF Purpose Mapping} \label{methodology_catego}
% \subsubsection{Approach}

\begin{table*}[h!]
    \centering
    \begin{tabular}{l|rrr|rr|r}
    \hline
    & \multicolumn{3}{c|}{\textbf{Statistical Methods}} & \multicolumn{2}{c|}{\textbf{AI-based Methods}} & \multirow{2}{*}{\textbf{Manual Mapping}} \\
    \cline{2-6}
    \textbf{Domain} & \textbf{spaCy} & \textbf{TheFuzz} & \textbf{\shortstack{Sentence \\ Transformers}} & \textbf{GPT-4.1} & \textbf{o3} & \\
    \hline
    reddit.com & 1,2,9,10,11 & 2,5,6,7,11 & 1,2,3,4,5,6,7 & 1,3,4,5,6,7,10 & 1,3,4,5,6,7,10 & 1,3,4,5,6,7,10\\
    google.com & 1,2,4,6,9,11 & 2,4,5,6,7,8,9,11 & 1,3,4,5,6,9 & 1,2,3,4,5,6,7,8,9,10,11 & 1,2,3,4,5,6,7,8,9,10,11 & 1,2,3,4,5,6,7,9,10,11\\
    x.com & 9,10 & 1,10 & 1,10 & 1,10 & 1,10 & 1,10\\
    linkedin.com & 2,9,10 & 3,10 & 10 & 1,2,3,4,7,9,10 & 1,4,7,9,10 & 1,3,4,10\\
    microsoft.com & 1,2,3,4,6,9,10,11 & 1,2,4,10 & 1,2,3,4 & 1,3,4,10 & 1,3,4,10 & 1,3,4,10\\
    bit.ly & 2,6,9,11 & 1,2,4,5 & 1,3,4 & 1,3,4,5,6 & 1,3,4,5,6,8,9,10 & 1,3,4,5,6\\
    adobe.com & 1,2,6,9,10,11 & 1,8 & 1,10 & 1,2,3,4,6,8,10,11 & 1,4,5,6,8,9,10 & 1,3,4,5,6,7,8,9,10\\
    f5.com & 2,9,11 & 8 & 5,6 & 5,6,8,9,10 & 1,5,6,8,9,10 & 1,3,4,5,6\\
    europa.eu & 1,2,6,9,11 & 1 & 1 & 1 & 1 & 1\\
    gravatar.com & 1,2,9,11 & 5,10 & 1 & 1,10 & 1,10 & 1,10\\
    cloudflare.com & 2,9,10,11 & - & 1 & 1,7,8,10 & 1,2,8,9,10 & 1,2,8,9\\
    themeisle.com & 2,4,9,10,11 & 3,4,10 & 1 & 1,2,10 & 1,3,4 & 1,3,4\\
    nytimes.com & 1,2,4,6,9,10,11 & 1,2,4 & 1,2,3,4 & 1,3,4,7,9,10 & 1,3,4,7,9,10 & 1,3,4,7,9,10\\
    forbes.com & 1,2,6,9,10,11 & 2,5,6 & 1,2 & 1,3,4,5,6,9,10 & 1,3,4,5,6,9,10 & 1,3,4,5,6,9,10\\
    zoom.us & 2,4,6,9,10,11 & 2,3,4,5,6 & 1,2,3,4,5,6 & 1,3,4,5,6,8,9,10 & 1,3,4,5,6,8,9 & 1,3,4,5,6,8,9\\
    \hline
    \end{tabular}
    \caption{Combined pre-study results grouped by Statistical and AI-based methods.}
    \label{tab:pre_study_combined}
\end{table*}

To determine which IAB TCF purposes are reflected in the extracted banner texts, we tested various statistical and AI-based methods. For the statistical methods, we explored three approaches, each implemented using a different Python library --- spaCy, TheFuzz, and SentenceTransformers --- to determine semantic similarity, perform keyword-based matching (including exact and close matches using both basic and regular expressions), and generate semantic representations of texts to calculate similarity scores, respectively~\cite{spacy101,thefuzz,sbert,sbertSTS}. 
Further details on how each library was used are provided in Appendix~\ref{appendix_E}.

For the AI-based methods, we used the OpenAI models GPT-4.1 and o3 provided by an AI toolbox\footnote{Access restricted proxy, provided by our institution < anonymised>.}. 
This allows us to compare the results from a general-purpose GPT model with those of a deep reasoning o-series model~\cite{openai_cookbook}. 
The used prompt included step-by-step instructions for processing the provided cookie banner texts and for IAB TCF purposes. 
The output included the assigned purposes, each with a brief explanation, facilitating a better understanding of the AI's decision-making for subsequent analysis and evaluation.
To determine the most accurate method for mapping the extracted banner text to the IAB TCF purposes, a pre-study compared the outputs of all methods against a manually annotated subset of 15 websites from the sample. 

Table~\ref{tab:pre_study_combined} lists the websites from the subset used in the pre-study, along with the mapping results of the statistical methods and AI-based methods, as well as the corresponding manual mapping.
For clarity, the purpose labels defined in Section~\ref{sec:cookiesAndTCF} were shortened within the table (e.g., 1 instead of P1).
Overall, AI-based methods are more closely aligned with manual mapping than statistical methods. 
We found that the AI-based method using OpenAI's o3 model yielded the most accurate results. 
Therefore, we applied it to the reduced sample of websites in the main study, comprising 181 websites.

% The results, which are presented in Table~\ref{tab:pre_study_combined} in the Appendix~\ref{appendix_F}, showed that the AI-based method using OpenAI's o3 model provided the most accurate results and was therefore used for the mapping in the main study, where it was applied to the reduced website sample of 181 websites. 

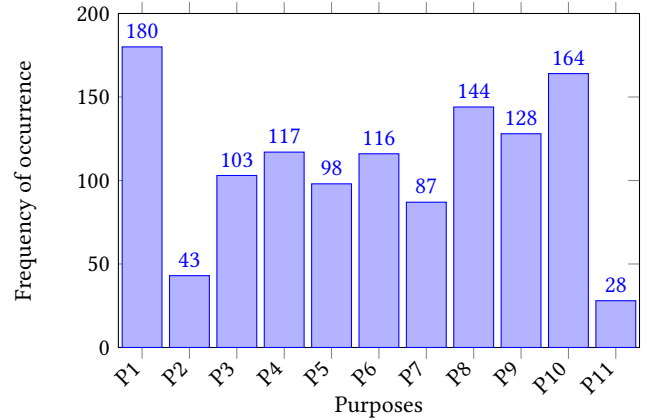
\begin{figure}[h!]
\centering
    \begin{tikzpicture}
      \begin{axis}[
        width=\linewidth,
        height=6cm,
        ybar,
        bar width=15pt,
        ylabel={Frequency of occurrence},
        xlabel={Purposes},
        symbolic x coords={P1,P2,P3,P4,P5,P6,P7,P8,P9,P10,P11},
        xtick=data,
        xticklabel style={rotate=45, anchor=east},
        enlarge x limits=0.05,
        ymin=0, ymax=200,
        every axis plot/.append style={fill=gray!40, draw=black},
        nodes near coords,
        nodes near coords align={vertical}
      ]
        \addplot coordinates {
          (P1, 180)
          (P2, 43)
          (P3, 103)
          (P4, 117)
          (P5, 98)
          (P6, 116)
          (P7, 87)
          (P8, 144)
          (P9, 128)
          (P10, 164)
          (P11, 28)
        };
      \end{axis}
    \end{tikzpicture}
\caption{Frequency of occurrence of purposes P1-P11 in the results of the main study.}
\label{fig:purposes-frequency}
\end{figure}

Figure~\ref{fig:purposes-frequency} illustrates how often each purpose was identified in the cookie banner texts of the reduced website sample.
\noindent As shown in Figure~\ref{fig:purposes-frequency}, the purposes P1 and P10 were assigned the most. In contrast, P2 and P11 were assigned the least. We observed that P1 was assigned whenever the cookie banner text mentioned the use of cookies. 
P2 was assigned when the banner text referred to advertising outside the context of personalisation. The AI model justified this assignment by assuming that such an advertisement might be selected given limited data. 
If the banner text referred to advertising in the context of personalisation, purposes P3 and P4 were generally assigned. Similarly, when the banner text referred to personalised content, purposes P5 and P6 were generally assigned. 
Although profiles were not explicitly mentioned, the AI model assumed their use whenever personalised advertising or content was discussed. We further observed that P3 and P4, as well as P5 and P6, were assigned together in most cases. The AI model justified this by reasoning that personalisation requires both the creation and usage of profiles. Therefore, P3 and P4 were generally assigned together when personalised advertising was mentioned, while P5 and P6 were generally assigned together when personalised content was mentioned. However, there were also cases in which only one of the paired purposes was assigned (e.g., P4 or P6), indicating an inconsistency in the AI model's assignment procedure. Purpose P7 was assigned when the banner text referred to measuring advertising performance. When content performance or traffic analysis was mentioned, purpose P8 was assigned. In some cases, where the banner text mentioned only performance metrics without explicitly referring to advertising or content, the purposes P7 and P8 were assigned, as they were interpreted in a broader context. Keywords such as~\enquote{audience insights} or~\enquote{preferences}, as well as the mentioning of sharing data with affiliates or partners, were mapped to P9. P10 was assigned when the banner text referred to product improvement or service development. P11 was assigned when the banner text referred to content outside the personalisation context. The AI model justified this assignment by assuming that such content might be chosen based on limited data.

There was no cookie banner text in the website sample that lacked a purpose assigned to it. Overall, some purposes, such as P1 and P10, are reflected in the banner texts through more explicit phrasing, whereas others, such as P2 and P11, are more implied than explicitly stated. The cookie banner texts of the reduced website sample revealed to contain vague wording (e.g.,~\enquote{assist in our marketing efforts},~\enquote{optimize your use of our website} or~\enquote{measurement of performance}) and diverse phrasings for the same underlying purpose (e.g.,~\enquote{understand users},~\enquote{audience insights} or~\enquote{audience research} for purpose P9). 

We found that the decision regarding the assignment of the purposes was often based on implications. The AI model made assumptions and subsequently drew conclusions about which purposes might be reflected in the cookie banner texts. As mentioned earlier, the use of profiles was rarely explicitly mentioned. However, the AI model considered them necessary to enable personalised advertising and content (i.e., the purposes P3–P6). The cookie banner texts also often included sections that described possible user actions (e.g., withdrawing consent) or subscription models (e.g., paying to avoid personalised ads). As these do not reflect any of the IAB TCF purposes, they were generally not assigned any purposes. In a few cases, all 11 purposes were assigned because the banner text directly referenced the IAB TCF and listed its purposes.

% 4.4. Modifying consent banner texts
\subsection{Modification of Cookie Banner Text} \label{modifiedBannerText}
    
To generate the modified cookie banner texts, a Python script was implemented that used the mapping results of the extracted cookie banner texts, along with the plain-language descriptions of the purposes. To generate the modified version of the cookie banner, the script maps each purpose identified during the mapping process to its corresponding~\plainLangDesc{}. In this way, we receive a modified version of the cookie banner text in which each purpose is accompanied by its~\plainLangDesc{}, replacing the original banner text.
% 5. Evaluation (Ref to sec. 2.3 and describe Table 5)
\section{Evaluation}
\label{sec:eval}

To evaluate the extent to which the cookie banner's textual content has become more understandable by the end of the process, comprehensibility is measured both before and after the modification. Table~\ref{tab:CBMB-cmprocess-excerpt} presents an excerpt of the results of the measurement of the comprehensibility of the cookie banner texts and the modified cookie banner texts. The full results can be found in the released artefact.
% Appendix~\ref{appendix_G}.

\begin{table}[h!]
\centering
\fontsize{6}{7}\selectfont
    \resizebox{\linewidth}{!}{%
    \begin{tabular}{p{1.8cm}cccccccccccc}
        \toprule
        & \multicolumn{2}{c}{\textbf{\shortstack{Flesch Reading\\Ease Score}}} 
        & \multicolumn{2}{c}{\textbf{\shortstack{FOG\\Index}}} 
        & \multicolumn{2}{c}{\textbf{\shortstack{Dale--Chall\\RS}}} 
        & \multicolumn{2}{c}{\textbf{\shortstack{capito.ai\\B1--B2}}} 
        & \multicolumn{2}{c}{\textbf{\shortstack{Wortliga\\Lang. Level}}} 
        & \multicolumn{2}{c}{\textbf{\shortstack{Wortliga\\CS}}} \\
        \cmidrule(lr){2-3}\cmidrule(lr){4-5}\cmidrule(lr){6-7}\cmidrule(lr){8-9}\cmidrule(lr){10-11}\cmidrule(lr){12-13}
        \textbf{Domain} 
        & {OB} & {MB}
        & {OB} & {MB}
        & {OB} & {MB}
        & {OB} & {MB}
        & {OB} & {MB}
        & {OB} & {MB} \\
        \midrule
        
        adobe.com      & 47.4  & 58.0  & 11.4  & 7.3  & 9.5  & 6.3  & 89\% & 94\% & B2 & B1 & 67 & 86\\
        bloomberg.com  & 35.17 & 66.44 & 12.68 & 7.09 & 8.52 & 6.24 & 11\% & 95\% & C2 & B1 & 26 & 86\\
        \textit{...}   & \textit{...} & \textit{...} & \textit{...} & \textit{...} & \textit{...} & \textit{...} & \textit{...} & \textit{...} & \textit{...} & \textit{...} & \textit{...} & \textit{...}\\
        \rowcolor{blue!20}
        \tikzmark{rowstart}mysql.com      & -56.43 & 66.54 & 47.92 & 7.11 & 14.43 & 6.12 & 0\% & 95\% & C2 & B1 & 0 & 86\tikzmark{rowend}\\
        tinyurl.com    & 36.18 & 66.44 & 12.04 & 7.09 & 8.86 & 6.24 & 14\% & 95\% & C2 & B1 & 47 & 86\\
        \textit{...}   & \textit{...} & \textit{...} & \textit{...} & \textit{...} & \textit{...} & \textit{...} & \textit{...} & \textit{...} & \textit{...} & \textit{...} & \textit{...} & \textit{...}\\
        \bottomrule
    \end{tabular}
    }

    \begin{tikzpicture}[remember picture, overlay]

      \draw[-, blue!70!black, thick] ([yshift=3pt,xshift=-5pt]pic cs:rowstart) -- ++(1,-1.35);
      \draw[-, blue!70!black, thick] ([yshift=3pt,xshift=-66pt]pic cs:rowend) -- ++(-1,-1.35);
    
    \end{tikzpicture}
    
    \vspace{1em}

    \fcolorbox{blue!50!black}{blue!3!white}{%
  \parbox{0.4\textwidth}{%
    \centering
    \fontsize{6}{7}\selectfont
    \textbf{Flesch Reading Ease Score} {↑} = increased readability \\[3pt]
    \textbf{FOG Index \& Dale–Chall RS} {↓} = lower grade level \\[3pt]
    \textbf{capito.ai \& Wortliga} {↑} = improved comprehensibility \& language level
  }%
}
\vspace*{0.3cm}
\caption{Excerpt of the comprehensibility measurement results for the cookie banner texts (OB) and their modified versions (MB). Abbreviations: RS = Readability Score, Lang. Level = Language Level, CS = Comprehensibility Score.}
\label{tab:CBMB-cmprocess-excerpt}
\end{table}

\noindent For the Flesch Reading Ease Score, the modified versions of the cookie banner texts in general received higher scores than their original versions, indicating an improvement in readability. In one case, the cookie banner text received a score of 8.74, while its modified version received a score of 57.98. In only 13 of 181 cases did the original cookie banner texts receive higher scores than their modified versions. For the FOG Index, the modified versions of the cookie banner texts generally received lower scores than their original versions. For example, the score of one cookie banner text decreased from 17.33 to 8.1 after it was modified. This means that the original version was understandable only to a college graduate, while the modified version is understandable to a seventh grader. In only 4 out of 181 cases, the original cookie banner texts received a lower score than their modified versions. For example, in two cases (6.52 vs.~8.53 and 8.77 vs.~8.85), the original versions received lower scores. However, the differences are minimal. A similar pattern can be observed with the Dale-Chall Readability Score: the modified versions of the cookie banner texts, in general, received lower scores than the original versions. For example, the score of one of the cookie banner texts decreased from 14.43 to 6.12 after being modified. This means that the original version was understandable for a college graduate, while the modified version is understandable for a 7th- or 8th-grade student. In 5 out of 181 cases, the original cookie banner texts received lower scores than their modified versions. For example, in two of these cases (9.02 vs.~9.24 and 6.75 vs.~7.27), the original version received a slightly lower score.
\newline
\newline
The percentages computed by capito.ai for the language levels B1–B2 showed higher values for the modified versions of the cookie banner texts in 167 out of 181 cases. In one case, the percentage even increased from 0\% to 97\%. Among the remaining cases, three showed no difference, while in 11 cases, the original cookie banner texts received higher percentages. For the modified versions, the percentage is at least 90\% in all cases. For the original versions, however, the percentage ranges from 0\% to 100\%. The language levels determined by the Wortliga text analysis tool generally showed improvement. In 142 of the 181 cases, the modified cookie banner text was assigned a lower language level than the original (e.g., C2 to B1 or C1 to B1). In 23 cases, the level remained the same (B1 or B2), and in the remaining 16 cases, the original cookie banner text had a higher language level than the modified version (e.g., A2 vs. B1, B1 vs. B2, or A1 vs. B2). The comprehensibility scores that were computed by the text analysis tool of Wortliga showed that the original cookie banner texts received scores ranging from 0 to 100, while the modified versions received scores ranging from 75 to 89. In 158 out of 181 cases, the comprehensibility scores improved. For example, the score increased from 0 to 87 in one case. There was also a case in which the score did not change. In the remaining 22 cases, the score decreased after the modification. For example, in one case, the score dropped from 99 to 82.
% 6. Discussion
\section{Discussion}
\label{sec:discussion}
This section discusses the main findings of our evaluation. We analyse the results of our mapping of cookie banner texts to the IAB TCF purposes, as well as the impact of applying plain language to improve comprehensibility. We also discuss the ethical considerations, limitations of our study, and ongoing GDPR discussions.

% 6.1. TFC Purpose cats
\subsection{IAB TCF Purpose Mapping}
As described in Section~\ref{methodology_catego}, the mapping process showed that vague wording and diverse phrasing for the same underlying purpose led the AI model to make assumptions about which purposes might be reflected in the banner text. This resulted in inconsistencies and generalisations during the assignment process, highlighting a limitation of automated purpose detection. These results indicate that ambiguous language and structural inconsistencies hinder both automated and human understanding. In addition, since some assignments were based on incomplete banner texts, as described in Section~\ref{cookieBTX}, not all purposes may have been identified, or in some cases, were even incorrectly assigned.

% 6.2. application of plain language
\subsection{Application of Plain Language}
\label{applicPL}
The application of plain language improved the readability of both the IAB's user-friendly descriptions and the cookie banner texts, as seen in Section~\ref{sec:AddressingCompr} and ~\ref{sec:eval}. The revised \plainLangDesc{}s achieved lower reading levels (B1–B2) and higher comprehensibility scores than the originals, which often required college-level reading ability. Similarly, the comprehensibility measurement showed that the modified cookie banner texts generally improved in readability. Regarding comprehensibility, we need to be aware that even if a text receives high scores indicating comprehensibility, this does not necessarily mean that its content is informative or helps users make an informed decision about the processing of their personal data. For example, one cookie banner that received a language level of A1 and a comprehensibility score of 100 contained only the following:~\enquote{This site uses cookies. By continuing to browse the site, you are agreeing to our use of cookies. Read our privacy policy.} Although this banner text received perfect scores for comprehensibility, it provided no specific information about how personal data is processed, aside from the fact that cookies are used on the site. This illustrates that plain language can make content more comprehensible, but cannot improve the informative nature of cookie banner content.

% 6.3. Limination (cut extension!) 
\subsection{Limitations}
\label{sec:limitations}
%\lex{we have no user study as oos and future work}
In this section, we outline the limitations regarding data collection, as well as those relating to linguistic, legal, and methodological aspects.

%\hd{I removed it, as we already mentioned it in the data collection}
% \paragraph{First Level of the Cookie Banner}
% We only considered the textual content on the first level of the cookie banner. As a result, we do not consider information contained in further levels, even though such information might also be relevant for users when making an informed decision regarding the processing of their personal data. This approach was chosen deliberately, as we want to focus on the part of the banner that most users actually interact with.
\paragraph{Variety of Cookie Banner Implementations} During our feasibility study, we came across various cookie banner variations, including elements such as dynamically generated structures, attributes like~\texttt{display} or accessibility-related attributes like~\texttt{aria-expanded}, and implementations using iFrames or Shadow DOM. We further observed that the cookie banner texts on the website sample used vague wording and varied phrasing for the same underlying purpose. This variety led the AI model to repeatedly make assumptions and draw conclusions about which purposes might be reflected in them. In some cases, this resulted in inconsistencies during the mapping process. The variety of cookie banner implementations, therefore, not only limited the extraction of textual content but also the proper assignment of purposes during the mapping process. Consequently, it limited our analysis of the current cookie banner landscape.

\paragraph{Manual Mapping as a Basis for Method Selection}
As described in Section~\ref{methodology_catego}, a pre-study was performed to determine the most accurate method for mapping the extracted banner texts to the IAB TCF purposes. Although the manual mapping (i.e., the assignment of IAB TCF purposes to the extracted banner texts from the subset of the website sample) was carried out to the best of our knowledge and based on careful interpretation of the purposes, it must be noted that the mapping is inherently subjective and may not be free of error. Because the manual mapping influenced the selection of the mapping method, this subjectivity must be considered when interpreting our findings.

\paragraph{Lack of Standardisation for Plain Language}
There are no fixed rules for plain language, as there are only recommendations for its use. Although the first DIN standards for plain language have been published (e.g., DIN ISO 24495-1 and DIN 8581-1), their use is only mandatory under certain conditions, such as when they are explicitly referenced in laws and regulations~\cite{DIN_BriefIntroductionToStandards}. As a result, even the language levels in which plain language is reflected are not clearly defined.  
Our research follows the definition of capito.ai, which associates plain language with the language levels B1-B2. However, other sources, for example, associate it with the language levels A2-B1~\cite{multisprech2024tools}. Beyond language levels, the application of plain language on cookie banner texts should also be discussed from a legal perspective to determine whether it can be used in practice in accordance with the law.
\paragraph{Limitations in Measuring Comprehensibility}
As described in Section~\ref{applicPL}, it must be emphasised that even if a text is considered comprehensible according to the measurement results, this does not imply that it contains relevant information that users need to make an informed decision about the processing of their personal data. Therefore, the actual effectiveness of comprehensibility is influenced by additional factors beyond the scope of our work. A user study should be conducted in which users interact with both the original cookie banner text and its modified versions to gain deeper insights into the cookie banner's comprehensibility.

\subsection{Ethical Consideration}\label{ethix}
% \hd{Todo: Responsible}
Our data collection was limited to the textual content of first-level cookie banners from publicly accessible websites. 
To ensure a non-invasive methodology, we did not bypass technical protections or access restricted content, and we excluded any websites where extraction was not readily feasible. This study involved no collection or processing of personal data, and the resulting dataset was used exclusively for research purposes.
%
% \lex{no invasive measurements (i.e., banner/website interaction), small dataset/no heavy load, no human involvement (i.e. user survey)}

% \subsection{Ongoing GDPR Discussion}
% \lex{something is going on in europe regarding change of GDPR. Is it relevant to us (anything there about ePrivacy)? if not - only the fact that a change of GDPR is in discussed again proofs that its not working right now!\\
% Or depend on the outcome, maybe it make sense to put in in discussion: even if our approach is not feasible YET, hopes are there GDPR/ePD gets more regulated in future.}
% 7. Conclusion
\section{Conclusion}
\label{sec:conclusion}

This study investigated the feasibility of utilising plain language (Einfache Sprache) to improve the comprehensibility of cookie banners within the IAB Transparency \& Consent Framework (TCF).
Our investigation of 181 websites revealed a cookie banner landscape dominated by technical jargon and vague phrasing, therefore reducing the comprehensibility of cookie banner texts.
Our results show that leveraging AI to map this complexity into clear, plain-language descriptions makes digital transparency in data processing achievable. We saw comprehension barriers fall from a college graduate level to that of a 7th-grade student.

At the same time, our findings suggest that the effectiveness of plain language is inherently limited by the quality of the original content; it cannot compensate for banners that fundamentally lack the necessary information.

Our work concludes that, while plain language is a vital tool for digital accessibility, standardised implementation guidelines for cookie banners are required to prevent complex or vague language being used as a `dark pattern'. This will ensure that all users can exercise their right to make informed decisions about their personal data.

\section*{ACKNOWLEDGMENTS}
In addition to the methods described in Section \ref{sec:method}, the authors used generative AI-based tools Grammarly and ChatGPT-5 to revise the text, correct any typos, grammatical errors, and awkward phrasing. 
No content was generated related to technical results, data, code, or analysis.

\bibliographystyle{ACM-Reference-Format}
\bibliography{ref_new}

% \cleardoublepage

\appendix
\onecolumn
\section{\textbf{USER-FRIENDLY} descriptions for the purposes defined by the IAB TCF} \label{appendix_A}

Table~\ref{tab:purpose-desc} lists the 11 standard purposes (P1–P11) defined by the IAB Transparency and Consent Framework (TCF) together with their explanations.

\begin{table}[h!]
% \small
     \centering
    
    \resizebox{\textwidth}{!}{
    \begin{tabular}
    {p{0.1\textwidth}|p{0.9\textwidth}}
    \toprule
    \textbf{Label} & \textbf{User-Friendly Description} \\
    \midrule
    P1 & Cookies, device or similar online identifiers (e.g. login-based identifiers, randomly assigned identifiers, network based identifiers) together with other information (e.g. browser type and information, language, screen size, supported technologies etc.) can be stored or read on your device to recognise it each time it connects to an app or to a website, for one or several of the purposes presented here.\\
    \hline
    P2 & Advertising presented to you on this service can be based on limited data, such as the website or app you are using, your non-precise location, your device type or which content you are (or have been) interacting with (for example, to limit the number of times an ad is presented to you).\\
    \hline
    P3 & Information about your activity on this service (such as forms you submit, content you look at) can be stored and combined with other information about you (for example, information from your previous activity on this service and other websites or apps) or similar users. This is then used to build or improve a profile about you (that might include possible interests and personal aspects). Your profile can be used (also later) to present advertising that appears more relevant based on your possible interests by this and other entities.\\
    \hline
    P4 & Advertising presented to you on this service can be based on your advertising profiles, which can reflect your activity on this service or other websites or apps (like the forms you submit, content you look at), possible interests and personal aspects.\\
    \hline
    P5 & Information about your activity on this service (for instance, forms you submit, non-advertising content you look at) can be stored and combined with other information about you (such as your previous activity on this service or other websites or apps) or similar users. This is then used to build or improve a profile about you (which might for example include possible interests and personal aspects). Your profile can be used (also later) to present content that appears more relevant based on your possible interests, such as by adapting the order in which content is shown to you, so that it is even easier for you to find content that matches your interests.\\
    \hline
    P6 & Content presented to you on this service can be based on your content personalisation profiles, which can reflect your activity on this or other services (for instance, the forms you submit, content you look at), possible interests and personal aspects. This can for example be used to adapt the order in which content is shown to you, so that it is even easier for you to find (non-advertising) content that matches your interests.\\
    \hline
    P7 & Information regarding which advertising is presented to you and how you interact with it can be used to determine how well an advert has worked for you or other users and whether the goals of the advertising were reached. For instance, whether you saw an ad, whether you clicked on it, whether it led you to buy a product or visit a website, etc. This is very helpful to understand the relevance of advertising campaigns.\\
    \hline
    P8 & Information regarding which content is presented to you and how you interact with it can be used to determine whether the (non-advertising) content, e.g. reached its intended audience and matched your interests. For instance, whether you read an article, watch a video, listen to a podcast or look at a product description, how long you spent on this service and the web pages you visit, etc. This is very helpful to understand the relevance of (non-advertising) content that is shown to you.\\
    \hline
    P9 & Reports can be generated based on the combination of data sets (like user profiles, statistics, market research, analytics data) regarding your interactions and those of other users with advertising or (non-advertising) content to identify common characteristics (for instance, to determine which target audiences are more receptive to an ad campaign or to certain content).\\
    \hline
    P10 & Information about your activity on this service, such as your interactions with ads or content, can be very helpful to improve products and services to build new ones based on user interactions, the type of audience, etc. This specific purpose does not include the development or improvement of user profiles and identifiers.\\
    \hline
    P11 & Content presented to you on this service can be based on limited data, such as the website or app you are using, your non-precise location, your device type, or the content you are (or have been) interacting with (for example, to limit the number of times a video or an article is presented to you).\\
    \bottomrule

    \end{tabular}
    }
    \caption{User-friendly descriptions for the purposes defined by the IAB TCF~\cite{IABTCF}.}\label{tab:purpose-desc} 
\end{table}

\clearpage

\section{Language levels as defined by capito.ai} \label{appendix_B}
\begin{figure}[h!]
        \centering
        \includegraphics[width=0.9\textwidth]{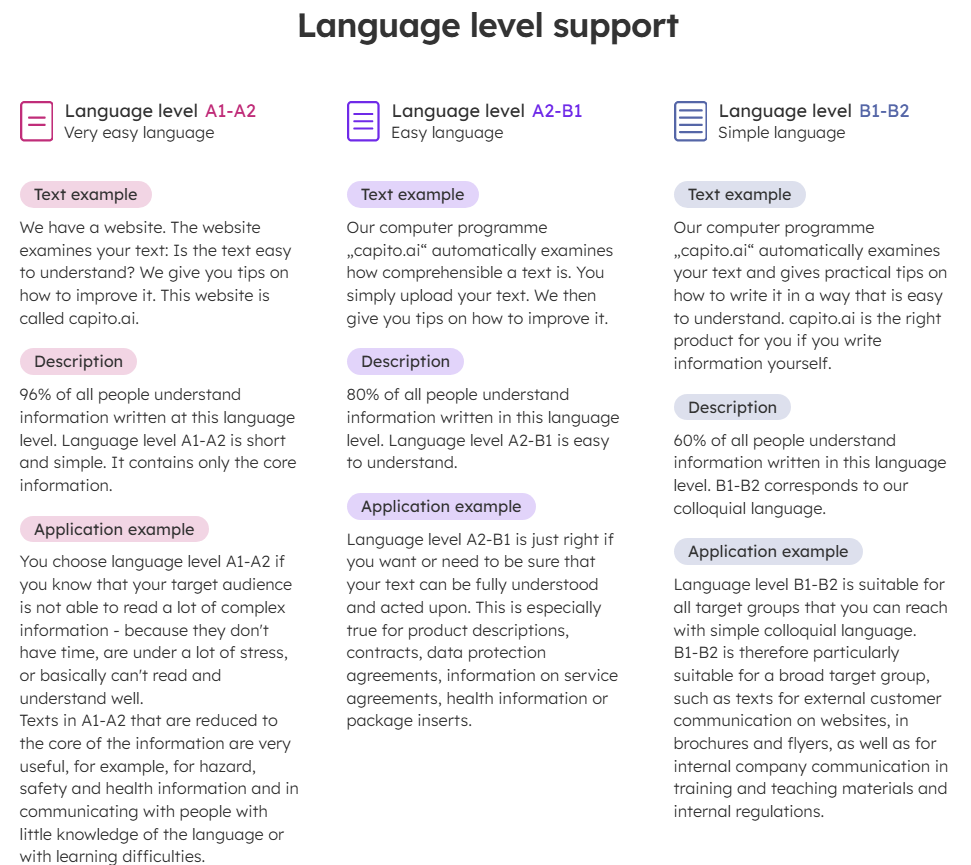}
        \caption{Language levels as defined by capito.ai~\cite{capitoAI}. In the German version of the page, \enquote{easy language} is referred to as Leichte Sprache and \enquote{simple language} as Einfache Sprache, which correspond to our definition of plain language as outlined in Section~\ref{sec:conceptPL}.}
        \label{fig:capito_screenshot}
\end{figure}

Figure~\ref{fig:capito_screenshot} shows three language levels defined by capito.ai: A1–A2 (very easy language), A2–B1 (easy language), and B1–B2 (simple language). For each level, examples, descriptions, and typical use cases are provided. 
The levels differ in complexity, target audience, and estimated comprehension rates, ranging from highly simplified core information (A1–A2) to standard simple language for broad audiences (B1–B2).

\section{Details on Statistical Methods} \label{appendix_E}

The open-source library spaCy is used for advanced natural language processing (NLP)~\cite{spacy101}. It offers a wide range of functionalities and allows us to determine the similarity between different texts by comparing their semantic representations. We use the TheFuzz Python library with keyword matching to detect not only exact matches but also close matches, such as slight variations of a word, when determining whether one of the purposes is included in the analysed cookie banner text~\cite{thefuzz}. For keyword matching, we define specific keywords for each purpose and consider them during the matching process. For the last method, we extend the keyword matching process by using regular expressions. Additionally, we use the SentenceTransformers Python module, also known as SBERT, to generate semantic representations of texts and calculate their similarity scores~\cite{sbert,sbertSTS}. %\hd{shall we put this one in the main text too?}

\end{document}